**Title:** Detecting Early Kidney Allograft Fibrosis with Multi-b-value Spectral Diffusion MRI

**Authors:** Mira M. Liu PhD[1], Jonathan Dyke PhD[2], Thomas Gladytz PhD[3], Jonas Jasse BS[4], Ian Bolger MS[1], Sergio Calle MS[1], Swathi Pavuluri BS[1], Tanner Crews BA[2], Surya Seshan MD[5], Steven Salvatore MD[5], Isaac Stillman MD[6], Thangamani Muthukumar MD[7], Bachir Taouli MD[1,8], Samira Farouk MD[9], Octavia Bane PhD[1,8,*], and Sara Lewis MD[1,8,*,*]

**Author Affiliations:**
[1]BioMedical Engineering and Imaging Institute, Icahn School of Medicine at Mount Sinai, New York, NY, USA.
[2]Department of Radiology/Citigroup Biomedical Imaging Center, Weill Cornell Medicine, New York, NY, USA.
[3]Berlin Ultrahigh Field Facility (B.U.F.F.), Max Delbrück Center for Molecular Medicine in the Helmholtz Association, Berlin, Germany.
[4]Department of Diagnostic and Interventional Radiology, Medical Faculty and University Hospital Düsseldorf, Heinrich-Heine-University Düsseldorf, Düsseldorf, Germany
[5]Department of Pathology, Weill Cornell Medicine, New York, NY, USA.
[6]Department of Pathology, Icahn School of Medicine at Mount Sinai, Mount Sinai Hospital, New York, NY, USA.
[7]Department of Nephrology and Kidney Transplantation Medicine, Weill Cornell Medicine, New York, NY, USA
[8]Department of Diagnostic, Molecular and Interventional Radiology, Icahn School of Medicine at Mount Sinai, Mount Sinai Hospital, New York, NY, USA.
[9]Transplant Nephrology, Icahn School of Medicine at Mount Sinai, Mount Sinai Hospital, New York, NY, USA.
[*] O. Bane and S. Lewis are co-PI senior authors

**\* Corresponding Author:**
Sara Lewis, MD
Professor of Diagnostic, Molecular and Interventional Radiology
BioMedical Engineering and Imaging Institute, Icahn School of Medicine at Mount Sinai
E-mail: Sara.Lewis@mountsinai.org

**Acknowledgements**
This project was funded by NIH NIDDK R01DK129888 (PI: Lewis/Bane) and supported by the National Center for Advancing Translational Sciences (NCATS) TL1TR004420 NRSA TL1 Training Core in Transdisciplinary Clinical and Translational Science (CTSA) (Fellow: Liu).

**ABSTRACT**
Kidney allograft fibrosis is a marker of chronic kidney disease (CKD) and predicts functional decline, and eventual allograft failure. This study evaluates if spectral diffusion MRI can help detect early development and mild/moderate fibrosis in kidney allografts. In a prospective two-center study of kidney allografts, interstitial fibrosis and tubular atrophy (IFTA) was scored and eGFR was calculated from serum creatinine. Multi-b-value DWI (bvalues = $[0,10,30,50,80,120,200,400,800 mm^2/s]$) was post-processed with spectral diffusion, intravoxel incoherent motion (IVIM), and apparent diffusion coefficient (ADC). Connection between imaging parameters and biological processes was measured by Mann-Whitney U-test and Spearman's rank; diagnostic ability was measured by five-fold cross-validation univariate and multi-variate logistic regression. Quality control analyses included volunteer MRI (n=4) and inter-observer analysis (n=19). 99 patients were included (50±13yo, 64M/35F, 39 IFTA=0, 22 IFTA=2, 20 IFTA=4, 18 IFTA=6, 46 eGFR≤45mL/min/1.73m$^2$, mean eGFR=47.5±21.3mL/min/1.73m$^2$). Spectral diffusion detected fibrosis (IFTA > 0) in patients with normal/stable eGFR >

45ml/min/1.73m$^2$ [AUC(95%$CI$) = 0.72(0.56, 0.87), p = 0.007]. Spectral diffusion detected mild/moderate fibrosis (IFTA=2-4) [AUC(95%CI) = 0.65(0.52, 0.71), p = 0.023], as did ADC [AUC(95%CI) = 0.71(0.54,0.87), p = 0.013)]. eGFR, time-from-transplant, and allograft size could not. Interobserver correlation was ≥0.50 in 24/40 diffusion parameters. Spectral diffusion MRI showed detection of mild/moderate fibrosis and fibrosis before decline in function. It is a promising method to detect early development of fibrosis and CKD before progression.
**Abbreviations and acronyms:** intravoxel incoherent motion (IVIM), interstitial fibrosis and tubular atrophy (IFTA), estimated glomerular filtration rate (eGFR), area under the curve (AUC), confidence interval (CI)

**INTRODUCTION**

Kidney allograft interstitial fibrosis and tubular atrophy (IFTA) is a marker of chronic kidney disease (CKD)[1] and associated with allograft failure and increased patient mortality[2,3]. While declining kidney function measured by an increase in serum creatinine in blood is a sign of CKD, the current reference standard of fibrosis is histopathology. This requires biopsy samples for diagnosis, staging of severity, continuous patient monitoring, as well as for studies of novel therapeutic outcome[4]. Further, fibrosis can develop silently before serum creatinine rises, and the time for intervention is lost as irreversible kidney damage occurs. As such, imaging may detect fibrosis without invasive biopsy, as well as provide information on kidney size, anatomy of the urinary system, and alternate diagnoses which biopsy and serum creatinine cannot. Noninvasive monitoring and subsequent early identification and quantification of fibrosis could enable therapeutic interventions that may preserve kidney function, including modifications in the patient's immunosuppressive regimen, and help screen for patients who may need further invasive biopsy.

Diffusion weighted magnetic resonance imaging (DWI) is a method of non-invasive measurement of kidney tissue diffusion characteristics without IV contrast, instead using diffusion weighting 'b-values'[5-10]. When a range of multiple b-values are used, the curve may diverge from a standard mono-exponential into a multi-exponential due to signal contribution from components with different diffusion coefficients. In the kidney, these components could include diffusion in the tissue parenchyma, kidney tubules, and capillary perfusion in vasculature. A such, multi-b-value DWI may add value to biopsy surveillance with whole kidney assessment of multiple physiologies to assess diagnosis, disease severity, and potential salvageability. Preliminary studies investigating multi-component model-free spectral diffusion in simulation[11-13], in healthy kidneys[14,15], and in native kidneys with CKD[16] suggest kidney allografts with reduced function and fibrosis may benefit from spectral diffusion that is sensitive to different physiologic components within a voxel.

As a step towards clinical translation, we evaluate multi-b-value MRI for the noninvasive diagnosis and quantification of fibrosis and function in kidney allografts in a prospective two-center study. We compare multi-component spectral diffusion that allows one to three components (vascular perfusion, tubular flow, and tissue parenchyma)[13,17] to two-component intravoxel incoherent motion (IVIM[18]; tissue component and vascular component[7,19-23]), and standard apparent diffusion coefficient (ADC). We then compare diagnostic ability of univariate and multiparametric logistic regression models built from these three diffusion models to those from standard clinical parameters to examine clinical relevance in early detection of fibrosis.

## RESULTS
### Patient Demographics and Clinical Characteristics

The demographics and clinical characteristics of all 99 patients (64M/35F, 50±13y) are included in Supplement A1. Comparisons between sites, and interobserver subset are included in Supplement A2. Four control-volunteers (1F/3M, 38.5±11.8y) were scanned at Site 1. Right native kidneys were chosen for analysis, to avoid tissue-air interface artifacts from the bowel, more prominent in the left kidney.

### Diffusion Spectra

Example DWI (b=0), T2w HASTE and corresponding example voxels with multi-b-value DWI curves and corresponding diffusion spectra are shown in Figure 1. Diffusion model parameters (Table 1), as well as example spectral diffusion parameter maps (Figure 2) are included in methods and materials.

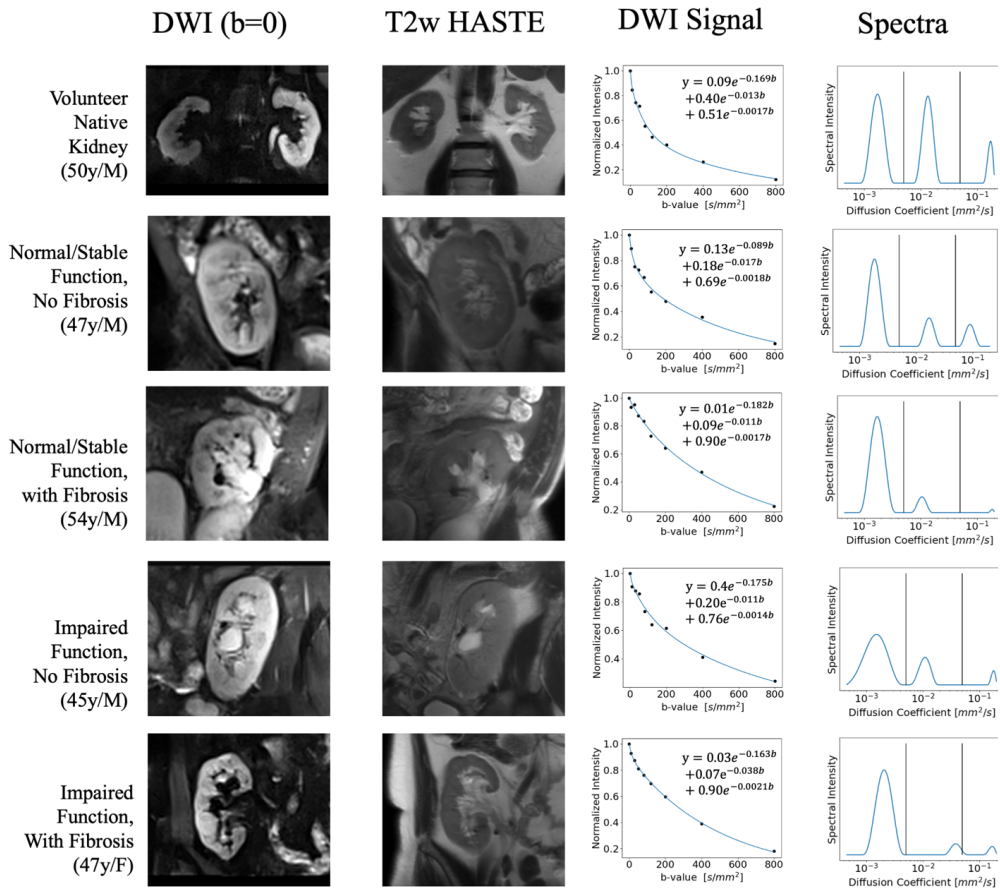

**Figure 1.** Example DWI and T2 weighted HASTE images of volunteer native kidneys and allografts for each of the four classifications of function and fibrosis, labeled for each row. An example multi-b value DWI curve from a voxel in each of the rows is shown in the third column. The corresponding diffusion spectrum is shown in the fourth column, with the multi-exponential fit resulting from the spectrum plotted on top of the DWI curve in the third column. Vertical lines are shown to represent the boundaries used to separate spectral peaks (Supplement E).

## Advanced Diffusion in Allografts Compared to Control Kidneys

The boxplot in Figure 3 shows $fD_{tubule}$ decreasing between control kidneys, stable kidney allografts, and diseased allografts; in comparison, IVIM $fD^*$, and ADC showed no significant correlation.

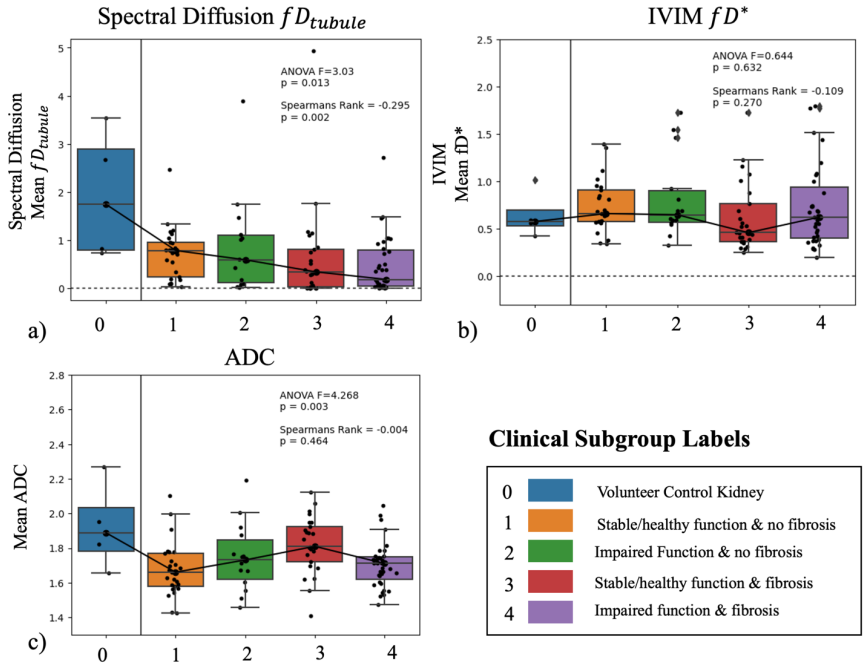

**Figure 3.** Box plots showing change in diffusion measured with **a)** spectral $fD_{tubule}$, **b)** IVIM $fD^*$, and **c)** ADC between volunteer native kidneys, and allografts with various renal functions and fibrosis scores. Kidneys were grouped ordinally by degree of renal disease as control volunteers (0), healthy allografts (1) to allografts with both impaired function and fibrosis (4) shown in the legend. 'Normal/stable function is determined as eGFR>45ml/min/1.73m², and 'fibrosis' determined as IFTA > 0. A line connecting the mean values for each type of kidney is plotted to show the trend. As fraction f is unitless, and diffusion coefficient D is in 10⁻³ mm²/s the units of $fD$ and $fD^*$ are 10⁻³ mm²/s and is a proxy for flow volume per unit time.

## Detection of Fibrosis

Diffusion model parameters with both significant Mann-Whitney U-test and area-under-the-curves (AUCs) for fibrosis are provided in Table 2. Spectral diffusion detected fibrosis in allografts (IFTA>0) with AUC(95%CI) = 0.69(0.59,0.81), p < 0.001 (Table 2a). Allografts with fibrosis had significantly increased $f_{tissue}$ and $fD_{tissue}$, with reduced tubule and vascular component parameters. Mean $fD_{tissue}$ [AUC(95%CI) = 0.66(0.55, 0.78), p = 0.005] returned the highest univariate AUC. IVIM std $D$ was significant, but neither IVIM nor ADC multiparametric AUCs were significant.

Spectral diffusion detected mild/moderate fibrosis in allografts (interstitial fibrosis and tubular atrophy score (IFTA) =0 vs. IFTA=1-4; Table 2b), with AUC(95%CI) = 0.65(0.52, 0.77), p = 0.02. Allografts with mild/moderate fibrosis had increased tissue component parameters and reduced tubule and vascular component parameters. Mean $fD_{tissue}$ [AUC(95%CI) = 0.67(0.55, 0.80), p = 0.004] again had the highest univariate AUC. While IVIM and ADC also returned significant features, IVIM and ADC multiparametric models were not significant.

Only spectral diffusion detected severe fibrosis (IFTA=0 vs. IFTA=5-6; Table 2c) with AUC(95%CI) = 0.68(0.52, 0.85), p = 0.026. IVIM and ADC returned no significant features, and no significant multiparametric models. Results for every parameter across all fibrosis diagnoses are included in Supplement B.

**Table 2.** Diagnostic performance of MR parameters for (a) no fibrosis (IFTA=0) vs. fibrosis (IFTA >0), (b) no fibrosis vs. mild/moderate fibrosis (IFTA=1-4), and (c) no fibrosis vs. severe fibrosis (IFTA=5-6). Included are parameters with both Mann-Whitney U-test p<0.05 and univariate logistic regression p<0.05. Presented is the mean±stdev, Mann-Whitney U-test p-value, and five-fold cross-validation ROC analysis. Numbers of cases per category are shown in corresponding parentheses by each group. As $f$ is unitless, and diffusion coefficient $D$ is in $10^{-3}$ mm$^2$/s the units of $fD$ are $10^{-3}$ mm$^2$/s and a proxy for flow volume per unit time.

| Table 2a) Parameter | No Fibrosis (39) $\mu \pm \sigma$ | Fibrosis (60) $\mu \pm \sigma$ | p-val | AUC (95%CI) | SN | SP | J-stat cutoff |
|---|---|---|---|---|---|---|---|
| **Spectral Diffusion** | | | | | | | |
| mean $fD_{tissue}$ | 1.59± 0.29 | 1.74± 0.29 | 0.003 | 0.66(0.55, 0.78) | 0.56 | 0.82 | 0.534 |
| std $f_{tissue}$ | 0.18±0.04 | 0.15±0.06 | 0.004 | 0.66(0.55, 0.77) | 0.64 | 0.65 | 0.500 |
| std $D_{tubule}$ | 12.48±5.65 | 9.05±5.09 | 0.004 | 0.66(0.55, 0.76) | 0.64 | 0.60 | 0.464 |
| mean $D_{tubule}$ | 7.58±4.82 | 5.41±4.66 | 0.010 | 0.64(0.53, 0.75) | 0.77 | 0.55 | 0.467 |
| std $D_{vasc}$ | 85.05±8.62 | 70.11±26.88 | 0.011 | 0.64(0.54, 0.75) | 0.85 | 0.52 | 0.468 |
| mean $f_{tubule}$ | 0.09±0.06 | 0.07±0.05 | 0.012 | 0.63(0.51, 0.74) | 0.46 | 0.78 | 0.503 |
| median $fD_{tissue}$ | 1.52±0.29 | 1.67±0.29 | 0.013 | 0.64(0.52, 0.75) | 0.67 | 0.62 | 0.495 |
| std $f_{tubule}$ | 0.14±0.05 | 0.11±0.06 | 0.018 | 0.63(0.52, 0.75) | 0.56 | 0.65 | 0.500 |
| mean $D_{vasc}$ | 89.31±35.1 | 68.89±46.48 | 0.022 | 0.62(0.51, 0.73) | 0.87 | 0.38 | 0.415 |
| mean $f_{tissue}$ | 0.75±0.07 | 0.79±0.09 | 0.027 | 0.62(0.51, 0.73) | 0.74 | 0.50 | 0.494 |
| Multiparametric | | | <0.001 | 0.69(0.59, 0.81) | 0.72 | 0.60 | 0.496 |
| **IVIM** | | | | | | | |
| std D | 0.25± 0.07 | 0.21± 0.06 | 0.009 | 0.64(0.53, 0.76) | 0.44 | 0.80 | 0.507 |
| Multiparametric | | | 0.33 | 0.44(0.32, 0.56) | 0.23 | 0.78 | 0.513 |
| Table 2b) Parameter | No Fibrosis (39) | Mild/Moderate Fibrosis (40) | p-val | AUC (95%CI) | SN | SP | J-stat Cutoff |
| **Spectral Diffusion** | | | | | | | |
| mean $fD_{tissue}$ | 1.59±0.29 | 1.74±0.23 | 0.004 | 0.67(0.55, 0.80) | 0.59 | 0.82 | 0.530 |
| std $D_{tubule}$ | 12.48±5.65 | 9.18±5.2 | 0.008 | 0.65(0.53, 0.77) | 0.74 | 0.55 | 0.431 |
| median $fD_{tissue}$ | 1.52±0.29 | 1.67±0.26 | 0.013 | 0.65(0.53, 0.77) | 0.74 | 0.55 | 0.431 |
| mean $f_{vasc}$ | 0.07±0.03 | 0.05±0.04 | 0.020 | 0.63(0.51, 0.76) | 0.72 | 0.68 | 0.499 |
| Multiparametric | | | 0.02 | 0.65(0.52, 0.77) | 0.67 | 0.70 | 0.511 |
| **IVIM** | | | | | | | |
| std $D$ | 0.25± 0.07 | 0.21± 0.06 | 0.007 | 0.66(0.54, 0.78) | 0.44 | 0.82 | 0.507 |
| mean $(1-f)D$ | 1.24± 0.14 | 1.3± 0.12 | 0.017 | 0.64(0.51, 0.76) | 0.74 | 0.53 | 0.493 |
| median $(1-f)D$ | 1.26± 0.14 | 1.32± 0.12 | 0.025 | 0.63(0.51, 0.76) | 0.69 | 0.55 | 0.496 |
| Multiparametric | | | 0.55 | 0.54(0.41, 0.67) | 0.77 | 0.38 | 0.485 |
| Table 2c) | No Fibrosis (39) | Severe Fibrosis (20) | | AUC (95%CI) | SN | SP | J-stat Cutoff |
| **Spectral Diffusion** | | | | | | | |
| std $f_{tissue}$ | 0.18±0.04 | 0.14±0.03 | 0.002 | 0.74(0.61, 0.87) | 0.54 | 0.90 | 0.501 |
| std $D_{vasc}$ | 85.05±8.62 | 67.13±24.95 | 0.003 | 0.72(0.56, 0.89) | 0.85 | 0.65 | 0.475 |
| std $f_{tubule}$ | 0.14±0.05 | 0.11±0.04 | 0.028 | 0.66(0.51, 0.80) | 0.54 | 0.75 | 0.501 |
| Multiparametric | | | 0.026 | 0.68(0.52, 0.85) | 0.87 | 0.50 | 0.466 |

**Detection of Fibrosis in Allografts with Normal/Stable Function**

Diffusion model parameters with both significant Mann-Whitney U-test and AUCs for detecting fibrosis in allografts with normal/stable function are shown in Table 3. Spectral diffusion detected fibrosis in allografts presenting with normal/stable function [AUC(95%CI) = 0.72(0.56,0.87), p < 0.01] (Table 3a). Median $fD_{tissue}$ [AUC(95%CI) = 0.70(0.55, 0.86), p = 0.006] had the highest univariate AUC for spectral diffusion. Both ADC and IVIM also showed an increase in the tissue component parameters in patients with fibrosis, but only ADC returned significant multi-parametric model (Table 3a).

Allografts with both impaired function and fibrosis showed increased tissue compartment heterogeneity and decreased tubule component parameters compared to healthy allografts (Table 3b). Spectral diffusion did not detect allografts with impaired function but no fibrosis; while stdev $D_{tissue}$ was significant, it did not pass multiple comparisons correction. Results for every parameter across all clinical subgroups are included in Supplement C.

**Table 3.** Diagnostic performance of MR parameters between allografts with normal/stable function (eGFR>45ml/min/1.73m$^2$) and no fibrosis (IFTA=0) versus (a) normal/stable function and fibrosis (IFTA>0), (b) impaired function and no fibrosis, and (c) impaired function (eGFR<45ml/min/1.73m$^2$) and fibrosis. Included are parameters with both Mann-Whitney U-test p<0.05 and univariate logistic regression p<0.05. Presented are mean±stdev, Mann-Whitney U-test p-value, and five-fold cross validation ROC analysis. Numbers of cases per category are shown in corresponding parentheses by each group. As fraction $f$ is unitless, and diffusion coefficient $D$ is in 10$^{-3}$ mm$^2$/s the units of $fD$ are 10$^{-3}$ mm$^2$/s and is a proxy for flow volume per unit time.

| Table 3a) | Normal/stable function, no fibrosis (25) | Normal/stable function & fibrosis (25) | p-val | AUC(95%CI) | SN | SP | J-stat Cutoff |
|---|---|---|---|---|---|---|---|
| **Spectral Diffusion** | | | | | | | |
| median $fD_{tissue}$ | 1.51±0.26 | 1.74±0.31 | 0.006 | 0.70(0.55, 0.86) | 0.76 | 0.68 | 0.469 |
| median $f_{tissue}$ | 0.76±0.08 | 0.84±0.10 | 0.010 | 0.70(0.56, 0.85) | 0.84 | 0.48 | 0.494 |
| std $D_{tubule}$ | 11.8±3.92 | 8.69±5.37 | 0.018 | 0.67(0.51, 0.82) | 0.76 | 0.56 | 0.433 |
| std $f_{vasc}$ | 0.09±0.03 | 0.07±0.05 | 0.025 | 0.67(0.51, 0.83) | 0.80 | 0.56 | 0.499 |
| Multiparametric | | | 0.007 | 0.72(0.56, 0.87) | 0.84 | 0.64 | 0.464 |
| **IVIM** | | | | | | | |
| std $D$ | 0.26± 0.06 | 0.2± 0.06 | 0.002 | 0.74(0.61, 0.88) | 0.52 | 0.80 | 0.506 |
| mean $(1-f)D$ | 1.23± 0.15 | 1.33± 0.14 | 0.014 | 0.68(0.52, 0.84) | 0.80 | 0.64 | 0.494 |
| median $(1-f)D$ | 1.25± 0.15 | 1.35± 0.13 | 0.013 | 0.69(0.54, 0.85) | 0.72 | 0.64 | 0.499 |
| Multiparametric | | | 0.35 | 0.58(0.41, 0.75) | 0.48 | 0.72 | 0.506 |
| **ADC** | | | | | | | |
| mean ADC | 1.68± 0.16 | 1.78± 0.18 | 0.025 | 0.69(0.54, 0.85) | 0.64 | 0.80 | 0.510 |
| median ADC | 1.66± 0.15 | 1.79± 0.18 | 0.011 | 0.69(0.54, 0.85) | 0.64 | 0.80 | 0.510 |
| std ADC | 0.33± 0.12 | 0.25± 0.08 | 0.017 | 0.67(0.51, 0.83) | 0.52 | 0.80 | 0.516 |
| Multiparametric | | | 0.013 | 0.71(0.54, 0.87) | 0.76 | 0.72 | 0.489 |
| Table 3b) | Normal/stable function, no fibrosis (25) | Impaired function & fibrosis (35) | p-val | AUC(95%CI) | SN | SP | J-stat Cutoff |
| **Spectral Diffusion** | | | | | | | |
| std $f_{tissue}$ | 0.18±0.04 | 0.15±0.03 | 0.004 | 0.71(0.57, 0.85) | 0.56 | 0.86 | 0.502 |
| std $D_{tissue}$ | 1.0±0.28 | 0.90±0.69 | 0.006 | 0.69(0.55, 0.83) | 0.76 | 0.54 | 0.490 |
| mean $f_{tubule}$ | 0.10±0.06 | 0.06±0.05 | 0.016 | 0.65(0.51, 0.79) | 0.52 | 0.74 | 0.501 |
| std $f_{tubule}$ | 0.14±0.05 | 0.10±0.05 | 0.023 | 0.66(0.52, 0.80) | 0.52 | 0.77 | 0.501 |
| Multiparametric | | | 0.021 | 0.67(0.52, 081) | 0.68 | 0.63 | 0.491 |
| **IVIM** | | | | | | | |
| std $D$ | 0.26± 0.06 | 0.22± 0.06 | 0.025 | 0.66(0.51, 0.80) | 0.52 | 0.77 | 0.503 |
| Multiparametric | | | 0.29 | 0.58(0.43, 0.73) | 0.56 | 0.63 | 0.500 |

**Multi-Component $fD$ Correlated with IFTA score**

Spectral diffusion $fD_{tissue}$ correlated positively with IFTA score in patients with normal/stable function (Spearman's rank $= 0.359, p < 0.01$; Figure 4a). $fD_{tubule}$ correlated negatively with IFTA score (Figure 4b) while $fD_{vasc}$ did not achieve statistical significance (Figure 4c). IVIM mean $(1-f)D$ and mean ADC, both alternate measures of diffusion in tissue parenchyma, also correlated positively with IFTA score (Figure 4d-e).

Significant correlation was also seen between IFTA and $fD$ across the entire patient cohort, i.e. not dichotomized by kidney function ($p = 0.005 - 0.045$). However, there was no significant correlation within the subset of patients presenting with impaired function ($p = 0.158 - 0.521$). Correlation was predominately in those presenting with normal/stable function.

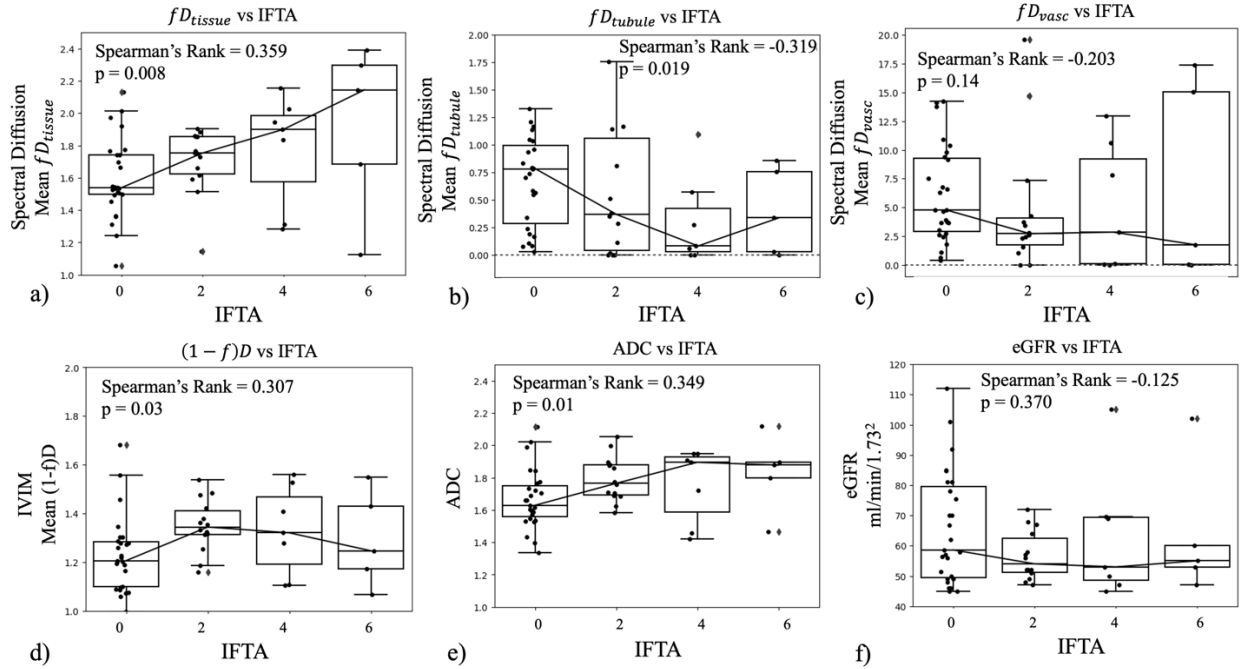

**Figure 4.** The top row shows spectral diffusion parameters of **a)** mean $fD_{tissue}$, **b)** mean $fD_{tubule}$, and **c)** mean $fD_{vasc}$ correlated against Banff 2017 IFTA scores in patients presenting with normal/stable eGFR> 45 ml/min/1.73m². The bottom row shows the same correlations for **d)** IVIM mean $(1-f)D$, e) mean ADC, and f) CKD-EPI eGFR.

**Diagnosis of Fibrosis with Clinical Parameters**

Estimated glomerular filtration rate (eGFR) detected fibrosis with AUC(95%CI) = 0.63(0.52,0.74), p = 0.027 (IFTA=0 vs. IFTA>0; Table 4). However, eGFR could not differentiate between no fibrosis and mild/moderate fibrosis or detect fibrosis if eGFR>45ml/min/1.73m2. Further, eGFR showed no correlation with fibrosis *within* the subsets of either normal/stable eGFR (Spearman's rank = −0.125, p = 0.370; Figure 4F) or impaired eGFR (Spearman's rank = −0.128, p = 0.520). Instead, eGFR differentiated between the severe fibrosis and mild/moderate fibrosis (Table 4) and correlated with IFTA score across *all* patients (Spearman's rank = −0.342, p < 0.01)

Allografts without fibrosis had shorter transplant intervals than those with severe fibrosis, i.e. an allograft was more likely to have developed fibrosis over time (Table 4). However, transplant interval was not significant for any other comparison in this study. Allograft volume, patient age, and BMI were not significant (p > 0.10) for any comparisons.

**Table 4.** Clinical demographic features eGFR, transplant-to-MRI interval (days), and allograft volume, for thresholding and grouping of IFTA. Included are parameters with both Mann-Whitney U-test p<0.05 and univariate logistic regression p<0.05. Presented is the mean±stdev, Mann-Whitney U-test p-value, and five-fold cross-validation ROC analysis. Numbers of cases per category are shown in corresponding parentheses by each group. No IFTA: ci+ct=0; mild/moderate IFTA: ci+ct= 1-4, severe IFTA: ci+ct= 5-6. Units of eGFR are in 45ml/min/1.73m².

| Table 4) | No Fibrosis (39) | Fibrosis (60) | p-val | AUC (95%CI) | SN | SP | J-stat Cutoff |
|---|---|---|---|---|---|---|---|
| eGFR | 54.08±23.67 | 43.16±18.36 | 0.014 | 0.63(0.52, 0.74) | 0.65 | 0.58 | 0.485 |

|  | No Fibrosis (39) $\mu \pm \sigma$ | Severe Fibrosis (20) $\mu \pm \sigma$ | p-val | AUC (95%CI) | SN | SP | J-stat Cutoff |
|---|---|---|---|---|---|---|---|
| eGFR | 54.08±23.67 | 36.3±19.81 | 0.003 | 0.72(0.58, 0.86) | 0.74 | 0.70 | 0.422 |
| Transplant Interval | 886±1679 | 2080±2454 | 0.015 | 0.71(0.55, 0.86) | 0.85 | 0.50 | 0.495 |
|  | Mild/Moderate Fibrosis (40) $\mu \pm \sigma$ | Severe Fibrosis (20) $\mu \pm \sigma$ | p-val | AUC (95%CI) | SN | SP | J-stat Cutoff |
| eGFR | 46.59±16.55 | 36.3±19.81 | 0.012 | 0.68(0.53, 0.83) | 0.65 | 0.75 | 0.489 |
| Transplant Interval | 1107±1783 | 2082±2454 | 0.016 | 0.63(0.48, 0.78) | 0.69 | 0.55 | 0.533 |

**Combined MR Diffusion and eGFR**

Combined spectral diffusion parameters and eGFR detected fibrosis (IFTA=0 vs IFTA>0), but it did not outperform spectral diffusion alone $[AUC(95\%) = 0.68(0.58, 0.79), p < 0.01; \text{DeLong } p = 0.58]$. Spectral diffusion alone had the highest AUC for detection of mild/moderate IFTA. Inclusion of eGFR, allograft volume, Transplant-to-MRI interval, patient age, and patient BMI decreased the mean AUC, which is expected as they were not significant.

**Interobserver Reproducibility and SNR**

Interobserver correlation ranged from poor to excellent (ICC range=0.03-0.92) with 5/40 features returning an ICC between 0 and 0.25, 11/40 features returning an ICC between 0.25 and 0.50, 14/40 with an ICC between 0.50 and 0.75, and 10/40 with ICC above 0.75. Tissue diffusion components showed better interobserver reliability than tubular components, and IVIM and ADC demonstrated better ICC and CoV(%) than spectral diffusion. A full table of all ICC and CoV% is included in Supplement D. The signal-to-noise ratio of the DWI b=0 kidney allografts, after motion correction and denoising, ranged from 30 to 50.

**DISCUSSION:**

Spectral diffusion was able to detect fibrosis and demonstrated good sensitivity for mild/moderate fibrosis that IVIM, ADC, eGFR, time-from-transplant, and allograft size did not. Further, it detected fibrosis in allografts that were still presenting with normal/stable eGFR. This could enable early detection of CKD before a decline in kidney function is evident, and inform decisions to pursue biopsy or change the medication regimen, as well as longitudinal monitoring in clinical trials of anti-fibrotic medications[24-26].

     This study expanded spectral diffusion from simulation[11] and control volunteers[14,15] to clinical translation of a multi-component diffusion model that includes more aspects of renal physiology. Unlike ADC[27], MR elastography[28], and IVIM[29], spectral diffusion separates diffusion components beyond vascular perfusion and tissue structure to provide insight into complex renal tubule physiology. Allografts showed lower tubular and tissue diffusion than volunteers, agreeing

with previously observed reduced diffusion and fluid transport[30]. Allografts with fibrosis also had lower vascular and tubular parameters which supports detecting damaged microvasculature and tubules in fibrotic and dysfunctional kidneys[31].

Across clinical comparisons, the tissue component showed the most clinical utility[15]. The fibrotic allografts had increased $fD_{tissue}$, supporting correlation with increased collagen deposition. While $D_{tissue}$ might be expected to decrease with fibrosis due to greater diffusion restriction from collagen, the increase in signal fraction of $f_{tissue}$ made the product $fD_{tissue}$ increase with fibrosis. This suggests $fD_{tissue}$ detects fibrous allograft tissue from the greater amount of the restricted diffusion, rather than slowed diffusion. Spectral diffusion improved diagnostic ability compared to IVIM $D$ (using a bi-exponential to remove fast diffusion contamination) or ADC with $b > 200$ (excluding low b-values dominated by fast diffusion signal). This supports advanced separation of diffusion components in kidney disease[32] to remove signal contamination from the tissue diffusion component and provide signal fraction.

Spectral diffusion separated mild/moderate fibrosis from no fibrosis while eGFR, allograft size, and time-to-transplant did not. Detection of mild/moderate fibrosis is clinically important as it may allow preventative intervention and treatment. Spectral diffusion could be another clinical measure, in addition to proteinuria, donor-derived cell-free DNA, and creatinine[33-35] for fibrosis detection and assessment of progression. Further, eGFR only significantly decreased once there were high levels of fibrosis, agreeing with current clinical knowledge; decreased eGFR tends to be a marker post the stage of irreversible fibrosis and considerable scarring[36]. In comparison, spectral diffusion detected fibrosis even when eGFR was normal/stable. This supports spectral diffusion detecting early change in microstructural diffusion patterns in the kidney and not being solely an indirect and more costly measure of GFR[11,13,15].

Results support $fD$ as a parameter of interest for multi-component flow in the kidney[13]. $fD$ of the vascular, tubule, and tissue components were significant and improved AUC values, as did IVIM $(1 - f)D$. $fD_{tubule}$ correlating negatively against IFTA scores supports detection of renal filtrate and tubule destruction in patients presenting with normal/stable function. Similarly, the positive correlation of $fD_{tissue}$ against IFTA scores supports detecting collagen deposition with nephron degeneration and tubular injury in allografts, rather lower ADC observed in native kidneys[37]. As there have been mixed results regarding reduced ADC detecting restricted diffusion in allografts, these results support $fD_{tissue}$ that includes the diffusion fraction may be a potential alternate along with previously observed corticomedullar difference[7,38,39].

Finally, spectral did not distinguish between fibrosis and no fibrosis for the subset with impaired function, nor between normal/stable function and impaired function for the subset with fibrosis. This highlights an important caveat: if a patient demonstrated impaired eGFR, spectral diffusion did not determine if the impaired allograft was fibrotic or not. However, detection of early fibrosis in patients presenting with normal/stable function remains clinically relevant.

We recognize several limitations. Further study is needed of spectral diffusion peak sorting, multi-component rigid models[8,40], and parameter stability. Whole cortex segmentation rather than circular ROIs may improve coverage and interobserver reliability, at the cost of artifacts. T2 effects, corticomedullary difference, and the influence of anisotropic collecting tubules in the medulla was beyond the scope of this work in allografts, but has shown promise in CKD of native kidneys[16]. Longitudinal study is needed to test if spectral diffusion can be an early predictor of fibrosis and function decline, and study of immune rejection in addition to fibrosis is warranted given that this is a potentially confounding pathologic variable[7]. While this study demonstrated potential clinical translation of $fD$, validation of $fD$ as a flow proxy may benefit from comparison to phantoms, complex simulations, microspheres, and flow cytometry in animal models[13,41], and radiotracers in human studies. This study included both protocol biopsies and clinically indicated biopsies which have some patient bias in terms of selection.

This work supports multi-component spectral diffusion detecting mild/moderate fibrosis development prior to kidney function decline, and correlation with fibrosis severity. Spectral diffusion could allow early intervention and preventative modification in medication regimen and treatment and reduce the need for repeated invasive biopsy or suggesting biopsy for potential intervention in patients still demonstrating healthy function. It could also contribute to longitudinal and novel therapeutic studies of fibrosis and CKD in kidney allografts.

**MATERIALS AND METHODS:**
**Patients**
This is a prospective, IRB-approved HIPAA-compliant two-center study at the Icahn school of Medicine at Mount Sinai and Weill Cornell Medicine that consists of kidney transplant recipients referred for percutaneous clinically indicated biopsies due to impaired allograft function or normal/stable function undergoing percutaneous protocol biopsies due to the presence of donor specific antibodies. Patients included in the study were those enrolled from 2/2022-09/2024 who are >1-month post-transplant. Informed consent was obtained, and patients underwent a non-contrast MRI protocol within 7 days of biopsy that included advanced DWI, as well as arterial spin labeling (ASL), blood oxygen level dependent imaging (BOLD), $T_1\rho$, $T_1$ relaxometry, and anatomical sequences (T2 HASTE, T1 in/opposed phase) that are beyond the scope of this current study. Exclusion criteria were age <18 years, large vessel or urinary tract complication of the kidney transplant, contra-indications to MRI, or pre-existing medical conditions including a likelihood of developing seizures or claustrophobic reactions. All experiments were performed in accordance with the Declaration of Helsinki, and allografts were procured according to the standard of care for subjects enrolled in the study at the Icahn school of Medicine at Mount Sinai and at Weill Cornell Medicine following relevant guidelines and regulations.

**Image Acquisition**
Patients underwent identical MR protocol with a 3T MRI (Mount Sinai: Skyra, Siemens Healthcare, Cornell: Prisma, Siemens Healthcare), set up by the same investigator at both sites, with a 16-channel body array and 32-channel spine array coils. The advanced DWI protocol was 2D coronal spoiled gradient echo-planar IVIM-DWI from the Siemens Advanced Body Diffusion works-in-progress package (WIP-990N) with respiratory gating (by liver-dome tracking, pencil-beam navigator). Averaged and motion-corrected trace-weighted DWIs were exported directly from the scanner with 'motion-corrected (MOCO)-averages', 'MOCO b-values', 'MOCO-3D', 'rescale local bias corruption' and denoising[42] selected for all 9 b-values (b-values= [0, 10, 30, 50, 80, 120, 200, 400, 800 $s/mm^2$]; TR/TE = 1500/58ms, voxel size = 2x2x5mm³, 4-directions, 16 slices, 3-averages, acquisition time ~7-15minutes). Control volunteers underwent the same protocol at Site 1.

**Image Analysis**

Six circular regions-of-interest (average ROI size: 64±25mm²) were delineated at the renal hilum on motion corrected b=0 s/mm² by a radiologist (Observer 1, 13 years of experience) using $T_2$-weighted images as reference (Horos v. 3.2.1, www.horosproject.org). Two ROIs were selected each at the upper pole, midpole, and lower pole, and propagated to each motion-corrected b-value. Voxel-wise analysis outperformed ROI-averaged signal and so is reported in this work.

**Spectral Diffusion Post-Processing**

The diffusion spectra of voxel-wise DWI decay curves within each ROI were calculated using non-negative least squares (NNLS) in MATLAB (Mathworks Inc, 2023b). The voxel-wise signal as a function of increasing b-value were fit to 300 logarithmically spaced D values ($\log_{10}(5)$-$\log_{10}(2200)$) as an unconstrained sum of exponentials (Eq. 1)[13][11,43].

$$y_i = \sum_{j=1}^{M} s_j e^{-b_i D_j} \quad (1)$$

In Eq. 1, $y_i$ is the equation for each of the N=9 b-values, for M=300 D values. $y_i$ as a function of b-value is the equation fit to the DWI decay curve. Minimizing the difference between Eq. 1 and the DWI decay curve, with Tikhonov regularization to smooth in the presence of noise, outputs a diffusion spectrum of the contributions of all 300 exponential basis vectors[11]. $\lambda$ was set at 0.1 to match optimal $\lambda \approx \frac{\#bval}{SNR}$[11] and reduce computation time.

$$\chi_r^2 = \min\left[\sum_{i=1}^{N}\left|\sum_{j}^{M} s_j e^{-b_i D_j} - y_i\right|^2 + \lambda \sum_{j=2}^{M-1}|s_{j+1} - 2s_j + s_{j-1}|^2\right] \quad (2)$$

The resulting spectra have peaks that represent the dominant basis vectors (Figure 1) per voxel without a priori assumption of number of peaks. Each peak returns a signal fraction $f$ and mean diffusion coefficient $D$, and spectral peaks can be sorted into (1) vascular, (2) tubular, and (3) tissue parenchyma components. A diffusion spectrum with three components would fit a tri-exponential equation as follows.

$$\frac{S_b}{S_{b0}} = f_{vasc}e^{-bD_{vasc}} + f_{tubule}e^{-bD_{tubule}} + f_{tissue}e^{-bD_{tissue}} \quad (3)$$

Voxels with $R^2 < 0.70$ were excluded from analysis. Example MR images with sample advanced DWI decay curve and spectral analysis are shown in Figure 1. Table 1 provides parameter definitions and the physiologic processes they may represent. Further detail regarding the fitting and analysis of diffusion spectra is included in Supplement E.

**IVIM and ADC Post-processing**

The voxel-wise DWI decay curve was fit to standard IVIM bi-exponential, $fe^{-bD^*} + (1-f)e^{-bD}$, with a Bayesian-log estimation[44] given priors log $D$ mean = 6.2±1 and log $D^*$ mean = 3.5±1[45]. A mono-exponential apparent diffusion coefficient (ADC) was calculated with a least-squares linear log fit of the signal from b = 200,400,800s/mm$^2$. This excluded IVIM effects at low b-values and non-Gaussian effects at high b-values[26]. Voxels with $R^2 < 0.70$ were excluded from analysis.

**Multi-component Diffusion $fD$ Parameter**

A parameter $fD$ was calculated for each diffusion component as the product of the fraction and diffusion coefficient of the individual spectral peaks as described by Liu et al[13] (Supplement E). In standard bi-exponential IVIM, $fD^*$ has been used as a marker of blood flow in a capillary network[19,20,46-48]. In this study, $fD$ was used as an estimate of local intravoxel 'flow' of every component[13]. $fD_{vasc}$ estimated the vascular motion, $fD_{tubule}$ estimated tubular motion, and $fD_{tissue}$ estimated total tissue diffusion in volume/time. For the standard bi-exponential, IVIM $fD^*$ and $(1-f)D$ were used to estimate the 'flow' of vascular and tissue components respectively. Figure 2 illustrates $fD$ of the three components.

**Table 1.** Spectral DWI parameters with corresponding hypothesized physiologic interpretation in the kidney cortex. Within the kidney cortex there are glomeruli, convoluted tubules, collecting ducts, and blood vessels. Conventional IVIM parameters are $f$ and $D^*$ for $f_{vasc}$ and $D_{vasc}$, and $D$ for $D_{tissue}$; there is no tubular component.

| Compartment | Spectral Parameters | Physiologic interpretation |
|---|---|---|
| Vascular: Vasculature in the kidney cortex is composed of blood vessels including arteries, veins, and capillaries. | $f_{vasc}$ | Fraction of a voxel that is within blood vessels |
| | $D_{vasc}$ | Diffusion coefficient, or speed, of the blood travelling within the vessels ($D^*$ in conventional IVIM) |
| | $fD_{vasc}$ | Proxy for the total blood flow in the vasculature. ($fD^*$ in conventional IVIM) |
| Tubular: Kidney tubules in the nephrons filter the glomerular filtrate and return nutrients to blood through reabsorption. The remaining fluid and waste become urine. In the cortex there are glomeruli and convoluted tubules. | $f_{tubule}$ | Fraction of a voxel that is within these tubules (no parameters in conventional IVIM) |
| | $D_{tubule}$ | Diffusion coefficient of the tubular filtrate (no parameters in conventional IVIM) |

|  | $fD_{tubule}$ | Proxy for the total tubular flow in the renal tubules (no parameters in conventional IVIM) |
| --- | --- | --- |
| Tissue Parenchyma: The kidney contains solid tissue, vascular endothelial cells, and tubular epithelial cells. Diffusion in the tissue parenchyma includes passive diffusion across cell membranes, and within the extracellular matrix (ECM) of the kidney parenchyma. Fibrosis in the kidney is a pathological feature that occurs when there is an excessive ECM accumulation leading to scarring and renal dysfunction. | $f_{tissue}$ | Fraction of a voxel that is composed of these solid cells in the kidney parenchyma with and without ECM accumulation. ($1 - f$ in IVIM, |
|  | $D_{tissue}$ | Diffusion coefficient of molecules within the parenchyma and composed of tissue both with and without ECM accumulation. |
|  | $fD_{tissue}$ | Proxy for the total restricted diffusive flow within solid tissue structure, cells, and in ECM scarring. |

**Interobserver Agreement**

Circular ROIs (average ROI size: 78±14mm$^2$) sampling the cortex of a subset of n=19 allografts, chosen from each clinical subgroup blinded to images, were delineated by an independent observer (Observer 2, a medical student with 1 year of experience) blind to original ROIs and diagnoses. Interobserver agreement was calculated via intraclass correlation coefficient (ICC) and coefficient of variation (CoV%) for all MR parameters. ROI placement and ROI size were not standardized between the two observers, but slice selection was held constant.

**Kidney Volume Measurement**

For assessment of three-dimensional volumetric measurement of the allograft in milliliters (ml), T1 in-phase images were copied to a post-processing workstation (Vitrea core, Vital Images, Minnetonka, MN, USA). Three-dimensional reconstruction was performed using semi-automated interposition based on signal intensity differences of the allograft compared to the surrounding tissues by Observer 1.

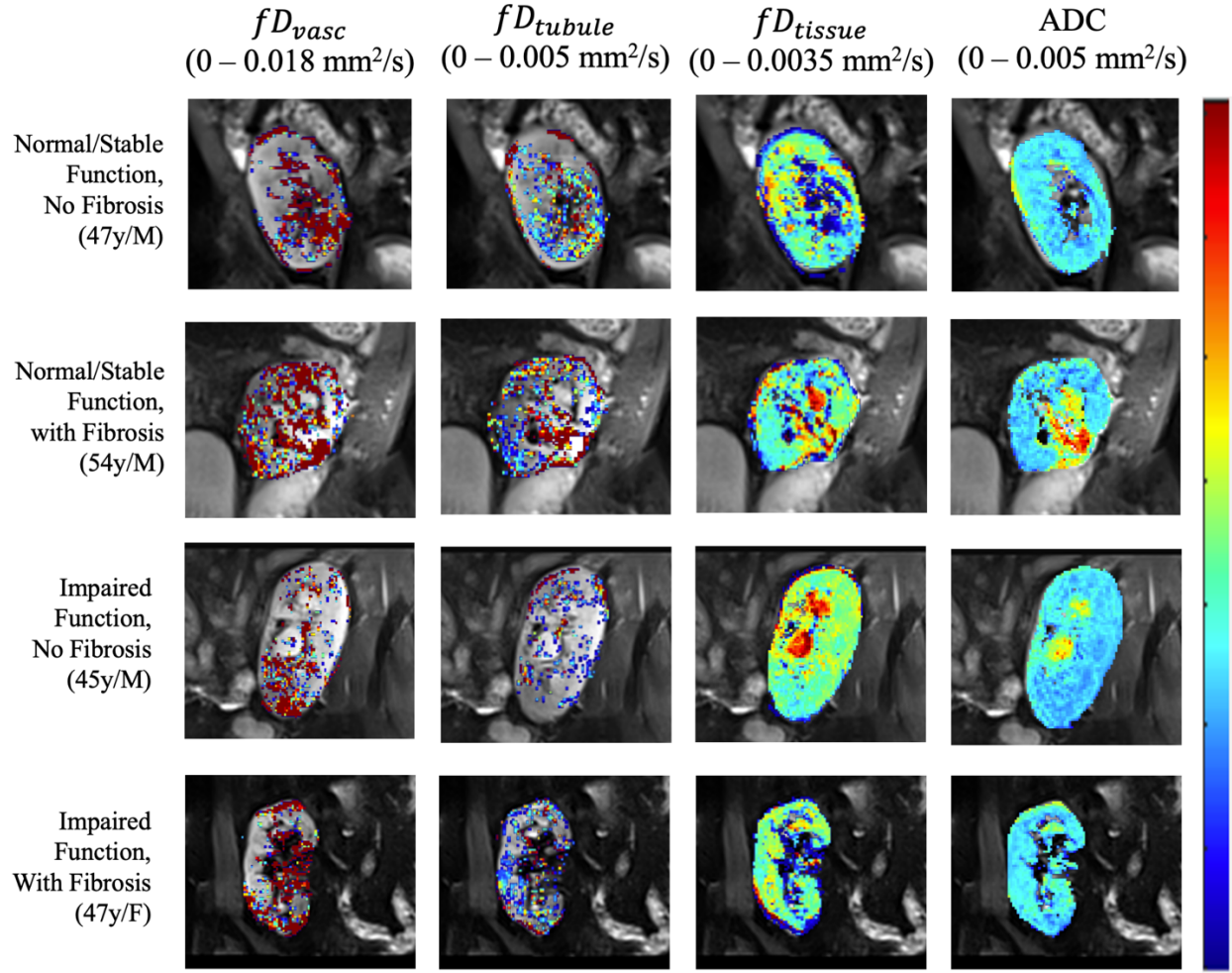

**Figure 2.** Images of different $fD$ components and ADC, superimposed on each respective b=0 DWI from Figure 1. These images are solely for illustration; they were not used for measurement. Unlike IVIM and ADC, spectral diffusion allows a flexible number of compartments and compartments that return 0 are indistinguishable from noise. As such, vascular and tubular images appear noisy. The trend of decreased tubular and vascular flow in diseased allografts can be seen. Note the difference in scale for the three compartments with $fD_{vasc}$ having the largest range, and $fD_{tissue}$ having the smallest.

**Laboratory Values and Histopathology**

Serum creatinine was collected at time of imaging or biopsy for measurement of eGFR calculated with race agnostic CKD-EPI 2021 criteria[49]. Interstitial fibrosis and tubular atrophy (IFTA=ci+ct) scores (range, 0-6) by pathologists were extracted from the clinical biopsy report, scored according to the Banff 2017 classification[50]. Other Banff diagnoses within the allograft specimens were also recorded[50] (Supplement F), but inflammation/rejection is beyond the scope of this study.

**Diagnostic Classifications and Clinical Subgroups**

IFTA score was used to diagnose fibrosis (no fibrosis: IFTA=0, fibrosis: IFTA>0), and fibrosis severity (mild/moderate: IFTA=1-4, severe: IFTA=5-6). Normal/stable allograft function was determined as eGFR $\geq$ 45ml/min/1.73m$^2$, and impaired function determined as eGFR <

45ml/min/1.73m$^2$. A threshold of 45ml/min/1.73m2 was used to compensate for single kidney filtration. Allografts were further divided into clinical subgroups: allografts with (1) normal/stable function and no fibrosis, (2) impaired function no fibrosis, (3) normal/stable function and fibrosis, (4) impaired function and fibrosis.

**Statistical Analysis and Machine Learning**

To examine direct connection between imaging parameters and biological processes, histogram characteristics voxel-wise mean, median, and standard deviation of the cortical ROIs were analyzed with respect to laboratory values and diagnoses. Central tendency measures (mean, median) of each component's $fD$ were included as MR parameters. Significant parameters were determined with Mann-Whitney U-test $p < 0.05$. The Benjamini-Hochberg procedure was applied for multiple comparisons corrections with a generous false discovery rate of 0.20, set to reduce false negatives in a novel method. Correlation of MR parameters against IFTA score was calculated with Spearman's rank and difference between clinical subgroups determined with ANOVA.

To examine diagnostic ability of imaging parameters and their direct relation to underlying physiology, univariate supervised machine learning logistic regression were built using significant histogram parameters with 5-fold cross validation. Diagnostic performance was assessed via receiver operating characteristic (ROC) and area-under-the-curve (AUC); mean AUC and 95% confidence interval (95%CI) was calculated via bootstrapping, and AUCs compared via the DeLong test. Sensitivity (SN), specificity (SP), and the optimal probability cutoff was calculated at the Youden's J-statistic (J-stat cutoff).

To compare overall diagnostic ability, one multiparametric model was built using parameters from each sequence (spectral diffusion, IVIM, and ADC) with 5-fold cross-validation for each diagnostic classification. Histogram characteristics for the multiparametric models were chosen as parameters that had $p < 0.05$ within training sets. This reduced data leakage and model overfitting in a small preliminary dataset. All statistical analysis and machine learning was performed in Python 3.11.4 (Anaconda Inc., 2024).

**Code Availability**

The code developed for spectral diffusion analysis in this work is available at https://github.com/miramliu/Spectral_Diffusion

**Supplement A1.** Patient demographics.

| Demographics and Clinical Features | |
|---|---|
| Site | 71 Site 1, 28 Site 2 |
| Biopsy Type | |
|   Indication Biopsy | N=71 |
|   Protocol Biopsy | N=28 |
| Sex(F/M) | 35/64 |
| Race | |
|   Black/African American | 46 |
|   White | 20 |
|   Asian | 8 |
|   Other/Unreported | 25 |
| Age (years; mean±std, range) | 50.1±13.1 (22-82) |
| Weight (kg; mean±std, range) | 79.1±16.9 (44.5-129.0) |
| BMI (kg/m$^2$, mean±std, range) | 27.4±4.7 (17.4-38.2) |
| Allograft Volume (mL, mean±std, range) | 238±72 (100-573) |
| Living Donor | N=43; |
|   Donor Age (years; mean±std, range) | 44.6±12.0 (21-68) |
| Deceased Donor | N=54; |
|   Donor Age (years; mean±std, range) | 37.25±12.6 (5-70)* |
|   KDPI | All KDPI < 85 |
| Time since transplant (months; mean±std, range) | 42.9±61.9 (1.3-266) |
| U-Protein (mg/24hr; mean±std, range) | 140.6±257.8 (1.0-1513), 39 unknown |
| Calcium Inhibitor status | 89 Tacrolimus, 4 Cyclosporine, 6 none |
| Donor Specific Antibodies status | 80 negative, 12 positive, 7 unknown |
| **eGFR (CKD-EPI 2021 mL/min/1.73m$^2$; mean±std, range)** | 47.5±21.3 (8.0-112.0) |
| eGFR≤ 45 (mL/min/1.73m$^2$; mean±std, range) | N=46; 29.9±8.4 (8.0-44.0) |
| eGFR> 45 (mL/min/1.73m$^2$; mean±std, range) | N=53; 62.7±16.9 (45.0-112.0) |
| **Interstitial Fibrosis/ Tubular Atrophy (IFTA)** | |
| IFTA = 0 | 39 |
| IFTA = 2 | 22 |
| IFTA = 4 | 20 |
| IFTA = 6 | 18 |
| **Categories** | |
| eGFR>45 & IFTA=0 | 25 |
| eGFR>45 & IFTA>0 | 25 |
| eGFR≤45 & IFTA=0 | 14 |
| eGFR≤45 & IFTA>0 | 35 |

*one pediatric donor

**Supplement A2.** Demographic features for the subset of 19 cases used in interobserver comparison.

| Demographics and Clinical Features | |
|---|---|
| Site | 14 Site 1, 5 Site 2 |
| Biopsy Type | |
|   Indication Biopsy | N=14 |
|   Protocol Biopsy | N=5 |
| Sex(F/M) | 8/11 |
| Age (years; mean±std, range) | 48.7±12.9 (25-69) |
| BMI (kg/m$^2$, mean±std, range) | 28.3±4.5 (20.7-38.2) |
| Living Donor | N=10; |
| Deceased Donor | N=9; All KDPI < 85 |
| Time since transplant (months; mean±std, range) | 53.2±62.7 (2.0-205) |
| U-Protein (mg/24hr; mean±std, range) | 140.6±257.8 (1.0-1513), 39 unknown |
| Calcium Inhibitor status | 18 Tacrolimus, 0 Cyclosporine, 1 none |
| Donor Specific Antibodies status | 17 negative, 1 positive, 1 unknown |
| **eGFR (CKD-EPI 2021 mL/min/1.73m$^2$; mean±std, range)** | 40.2±18.0 (14.0-85.0) |
| eGFR≤ 45 (mL/min/1.73m$^2$; mean±std, range) | N=12; 29.4±10.4 (14.0-42.0) |
| eGFR> 45 (mL/min/1.73m$^2$; mean±std, range) | N=7; 58.5±12.7 (45.0-85.0) |
| **Interstitial Fibrosis/ Tubular Atrophy (IFTA)** | |
| IFTA = 0 | 8 |
| IFTA = 2 | 4 |
| IFTA = 4 | 2 |
| IFTA = 6 | 5 |
| **Categories** | |
| eGFR>45 & IFTA=0 | 4 |
| eGFR>45 & IFTA>0 | 3 |
| eGFR≤45 & IFTA=0 | 4 |
| eGFR≤45 & IFTA>0 | 8 |

Site 1 patients were recruited from kidney transplant recipients with either acute or chronic allograft injury undergoing percutaneous clinically indicated biopsies, while Site 2 patients recruited kidney transplant recipients undergoing percutaneous clinically indicated biopsies and patients undergoing surveillance biopsy in the setting of donor specific antibody positivity. The decision to perform percutaneous allograft biopsy was based on the clinician's adjudication of each patient's clinical presentation regarding institute specific management or prognostication. Mean age and gender distribution ($\chi^2 = 1.25, p = 0.263$) were not significantly different between the two sites though Site 2 recruited patients with a statistically higher average BMI. As Site 1 and Site 2 recruited different subsets of kidney transplant recipients, in line with expectations Site 2 had a greater number of low IFTA scored patients, and a higher average eGFR due to the surveillance biopsies.

**Supplement B.** Supplementary Tables: Fibrosis Scores

Tabular Data of Spectral Diffusion, IVIM, and ADC values between (a) no fibrosis (IFTA=0) vs. fibrosis (IFTA >0), (b) no fibrosis vs. mild/moderate fibrosis (IFTA=1-4), and (c) no fibrosis vs. severe fibrosis (IFTA=5-6). Features that returned a raw p<0.05 are colored red. Features that passed the Benjamini-Hochberg correction with fpr=0.20 are highlighted yellow.

**Supplement C.** Supplementary Tables: AllograftGroups

Tabular Data of Spectral Diffusion, IVIM, and ADC values between allografts with normal/stable function (eGFR≥45ml/min/1.73m$^2$) and no fibrosis (IFTA=0) versus (a) normal/stable function and fibrosis (IFTA>0), (b) impaired function and no fibrosis, and (c) impaired function (eGFR<45ml/min/1.73m$^2$) and fibrosis. Features that returned a raw p<0.05 are colored red. Features that passed the Benjamini-Hochberg correction with fpr=0.20 are highlighted yellow

**Supplement D.** Supplementary Tables: InterobserverRepeatability

Tabular Data of interclass correlation coefficient and corresponding F-stat and 95% CI for all parameters. The coefficient of variability is also provided. There is a low ICC for median fD tubule (ICC=0.03), due to low agreement of median f tubule ICC (0.04), as many f tubules are zero.

| Interobserver Repeatability | | | | |
|---|---|---|---|---|
| Feature | ICC value | F-stat | 95% CI | CoV% value |
| spectral mean f vascular | 0.75 | 7.1 | [0.46 0.9 ] | 25.914 |
| spectral median f vascular | 0.78 | 8.29 | [0.52 0.91] | 62.053 |
| spectral std f vascular | 0.44 | 2.57 | [-0.  0.74] | 22.751 |
| spectral mean f tubule | 0.35 | 2.1 | [-0.11 0.69] | 40.850 |
| spectral median f tubule | 0.04 | 1.09 | [-0.41 0.48] | 75.356 |
| spectral std f tubule | 0.55 | 3.42 | [0.14 0.8 ] | 30.253 |
| spectral mean f tissue | 0.47 | 2.78 | [0.03 0.76] | 5.148 |
| spectral median f tissue | 0.35 | 2.06 | [-0.12 0.68] | 4.457 |
| spectral std f tissue | 0.54 | 3.38 | [0.13 0.8 ] | 16.285 |
| spectral mean D vascular | 0.64 | 4.53 | [0.27 0.84] | 18.854 |
| spectral median D vascular | 0.38 | 2.22 | [-0.08 0.7 ] | 56.568 |
| spectral std D vascular | 0.11 | 1.24 | [-0.35 0.53] | 7.396 |
| spectral mean D tubule | 0.58 | 3.81 | [0.19 0.82] | 23.631 |
| spectral median D tubule | 0.4 | 2.34 | [-0.05 0.72] | 65.052 |
| spectral std D tubule | 0.76 | 7.3 | [0.48 0.9 ] | 18.990 |
| spectral mean D tissue | 0.21 | 1.55 | [-0.25 0.6 ] | 7.001 |
| spectral median D tissue | 0.45 | 2.65 | [0.01 0.75] | 5.724 |
| spectral std D tissue | 0.3 | 1.87 | [-0.16 0.66] | 15.868 |
| spectral mean fD vascular | 0.42 | 2.46 | [-0.03 0.73] | 9.325 |
| spectral mean fD tubule | 0.25 | 1.68 | [-0.21 0.63] | 50.211 |
| spectral mean fD tissue | 0.71 | 5.95 | [0.39 0.88] | 65.881 |
| spectral median fD vascular | 0.4 | 2.32 | [-0.06 0.72] | 10.330 |
| spectral median fD tubule | 0.03 | 1.06 | [-0.42 0.47] | 79.441 |
| spectral median fD tissue | 0.79 | 8.64 | [0.54 0.91] | 34.441 |
| IVIM mean D | 0.76 | 7.39 | [0.48 0.9 ] | 2.237 |
| IVIM median D | 0.67 | 5.12 | [0.33 0.86] | 2.421 |
| IVIM std D | 0.89 | 16.89 | [0.73 0.96] | 6.088 |
| IVIM mean D* | 0.81 | 9.48 | [0.57 0.92] | 11.101 |
| IVIM median D* | 0.55 | 3.43 | [0.14 0.8 ] | 10.649 |
| IVIM std D* | 0.92 | 23.97 | [0.8 0.97] | 15.268 |
| IVIM mean f | 0.66 | 4.82 | [0.3 0.85] | 5.694 |
| IVIM median f | 0.63 | 4.38 | [0.26 0.84] | 5.517 |
| IVIM std f | 0.32 | 1.92 | [-0.15 0.67] | 14.865 |
| IVIM mean (1-f)D | 0.78 | 8.2 | [0.52 0.91] | 3.114 |
| IVIM median (1-f)D | 0.71 | 6.01 | [0.4 0.88] | 3.140 |
| IVIM mean fD* | 0.73 | 6.52 | [0.43 0.89] | 13.477 |
| IVIM median fD* | 0.56 | 3.58 | [0.16 0.81] | 11.270 |
| mean ADC | 0.84 | 11.3 | [0.63 0.93] | 2.868 |
| median ADC | 0.79 | 8.74 | [0.54 0.92] | 3.069 |
| std ADC | 0.62 | 4.26 | [0.24 0.83] | 11.220 |

**Supplement E. Spectral Diffusion and Multi-Component Diffusion $fD$**

Spectral peaks were sorted into (1) vascular, (2) tubular, and (3) tissue parenchyma components based on diffusion coefficient. A spectrum with peaks in each of the three components listed above would be a tri-exponential, as follows.

$$f_{vasc}e^{-bD_{vasc}} + f_{tubule}e^{-bD_{tubule}} + f_{tissue}e^{-bD_{tissue}}$$

In this work, these three components are ordered from left to right from fastest to slowest diffusion coefficient. The peaks were sorted as follows: the largest peak closest to $1.8 \times 10^{-3}$ mm²/s, a tissue diffusion literature value between two-component and three-component rigid fits[8], was considered the tissue parenchyma peak if below $50 \times 10^{-3}$ mm²/s. Beyond that, peaks were sorted as $0.8 \leq$ tissue $< 5 \leq$ tubule $< 50 \leq$ vascular for diffusion coefficients in units of $10^{-3}$ mm²/s. The boundary of $5 \times 10^{-3}$ mm²/s for tissue parenchyma is higher than $2 \times 10^{-3}$ mm²/s to include the boundary of the edge of the tissue peak, even though the peak's maximum is closer to $2 \times 10^{-3}$ mm²/s (Figure 1). If more than one peak fell within a certain range, the peaks were combined weighted by their respective volume fractions for statistical analysis of diagnostic ability. Peaks with mean D<0.8 were excluded due to b=800 being too low to capture very slow diffusion reliably[11].

$fD$ of each peak was treated as a multi-component flow proxy as described in Liu et al[13]. With a Gaussian approximation of each physiologic component[48], $fD$ was used to estimate the amount of flow of any distinct diffusion component, without capillary network assumptions of conventional IVIM perfusion. Further, as the signal labeling and readout are in the same plane[51,52], $fD$ calculates local flow without needing an AIF[53-56]. Therefore, $fD$ may provide information on local tissue diffusion, tubular flow, and perfusion without the complexities of selecting an aortic AIF[57], a labeling-plane for ASL, or correction for delayed and dispersed contrast a local-AIF[58]. Further, non-vascular flow that may not have an AIF, such as diffusion in the renal tubules, can still be imaged with spectral diffusion.

**Supplement F. Clinical notes**

Table A1. Further clinical notes and diagnoses for the 14 patients with allografts that showed impaired function but no fibrosis in this preliminary study.

| | |
|---|---|
| Patient 1 | Active antibody mediated rejection, Chronic active antibody mediated rejection, Chronic active T-cell mediated rejection Grade II, Suspicious or borderline for acute T-cell mediated rejection, Transplant glomerulopathy, moderate glomerulitis, multilayering of interstitial capillary basement membranes, and focally positive (2+) C4d staining of interstitial capillaries, Arteriolar hyalinosis. protein droplets in podocytes positive for albumin and kappa |
| Patient 2 | Active antibody mediated rejection, Acute/active T-cell mediated rejection Grade IB, Acute/active T-cell mediated rejection Grade III, Acute/active vascular rejection Grade III, Recurrent disease, Focal segmental glomerulosclerosis, NOS. |
| Patient 3 | C4d deposition without evidence of rejection, Acute tubular injury, Focal crystals consistent with calcium phosphate, Mild arteriosclerosis. There is no significant tubular or interstitial staining for IgG, IgA, IgM, C3, C1q, Kappa and Lambda light chains, albumin, and fibrinogen. |
| Patient 4 | Acute tubular injury, Thin glomerular basement membranes |
| Patient 5 | Acute tubular injury. There is no significant glomerular, tubular, or interstitial staining for IgG, IgA, IgM, C3, C1q, Kappa and Lambda light chains, albumin, and fibrinogen. |
| Patient 6 | Mild glomerulitis, Mild interstitial lymphoid infiltrates. a scarred region of the glomerulus shows nonspecific smudgy staining for IgM, C3 and C1q |
| Patient 7 | None |
| Patient 8 | 1+ C3 seen in vessels, Polyomavirus Nephritis (01_64) |
| Patient 9 | Mild acute tubular injury |
| Patient 10 | negative CD4 |
| Patient 11 | None |
| Patient 12 | Mild glomerulitis |
| Patient 13 | None |
| Patient 14 | Mild glomerulitis |